\documentclass[12pt,showpacs,preprintnumbers,aps]{revtex4}
\usepackage{epsf}
\usepackage{dcolumn}
\usepackage{bm}
\draft

\begin{document}
\title{The phenomenon of time-reversal violating generation of static magnetic and electric fields
is a basis of a new method for measurement of the electron EDM and T-odd P-odd constants of electron
interaction with a nucleon beyond Standard Model}
\author{ V. G. Baryshevsky}
\affiliation{Research Institute for Nuclear Problems, Belarusian State University,
11 Bobryiskaya str., 220050, Minsk, Republic of Belarus,\\
E-mail: bar@inp.minsk.by }
\date{\today}

\begin{abstract}
%to change ???????????
It is shown that in the experiments for search of EDM of an electron (atom, molecule)   
the T-odd magnetic moment induced by an electric field and the T-odd electric dipole moment 
induced by a magnetic field will be also measured.
It is discussed how to distinguish these contributions.
\end{abstract}


%%%%%%%%%%%%%%%%%%%%%
\maketitle

\narrowtext

Nowadays there is an appreciable progress in development of methods for
measurement of ultra weak magnetic and electric fields. 
{Therefore, 
new experiments for measurement of an electric dipole moment(EDM) $d$ of electrons 
\cite{1} are being prepared}. 

The EDM of a particle exists if parity (P) and time-reversal (T) invariance
are violated. Investigation of the EDM existence could provide knowledge about
physics beyond the Standard Model [1-4].

F.L.Shapiro's idea \cite{5} to measure the electron EDM  by applying a strong electric field 
to a substance that has an unpaired electron spin is being used for the EDM search 
\cite{1,6}.

The interaction $W_E$ of the electron electric dipole moment $\vec{d}$  
with an electric field $\vec{E}$ depends on their orientation:
\begin{equation}
W_E=-\vec{d}\vec{E},
\label{1}
\end{equation}
where $\vec{d}=d \frac{\vec{J}}{J}$, $\vec{J}$ is the atom spin, 
d is the EDM.

\begin{figure}[htbp]
\epsfysize = 4 cm \centerline{\epsfbox{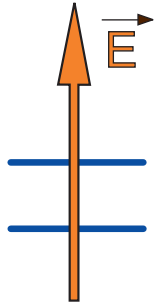}}
\caption{Splitting of levels in an electric field}
\end{figure}

Spins of electrons (atoms) at low temperature  appear to be polarized due to (\ref{1}) 
similar to the polarization (magnetization) of electrons by a magnetic field 
in paramagnetic substances 
due to the interaction $W_B$ of an electron (atom) magnetic moment $\vec{\mu}$ 
with a magnetic field $\vec{B}$
\begin{equation}
W_B=-\vec{\mu}\vec{B}.
\label{2}
\end{equation}
Spins of electrons (atoms) polarized by an electric field induce the magnetic
field $\vec{B}_E$ (Fig.2) and change in the magnetic flux $\Phi$ at the
surface of a flat sheet of material \cite{1}:
\begin{eqnarray}
\Delta \Phi=4 \pi \chi A d E^{*}/\mu_a,
\label{3}\\
{B_E}=\frac{\Delta \Phi}{A}=4 \pi \chi \frac{d}{\mu_a}{E^{*}},
\label{4}
\end{eqnarray}
where $\chi$ is the magnetic susceptibility, $\chi \approx \frac{\rho {\mu_a}^2}{3 k_{B} T}$,
$\rho$ is the number density of spins of interest, $k_B$ is Boltzmann's constant and $T$
is the sample temperature. In the cases where simple Langevin
paramagnetism is applicable, $E^{*}$ is the effective electric field at the location of
the spins, $\mu_{a}=g \sqrt{J(J+1)}\mu_{B}$ where $\mu_{B}$ is the Bohr magneton, 
$\mu_{a}$ is the atomic or ionic magnetic moment, 
$g$ is the Lande factor and $A$ is the sample area.
\begin{figure}[htbp]
\epsfysize = 3.5 cm \centerline{\epsfbox{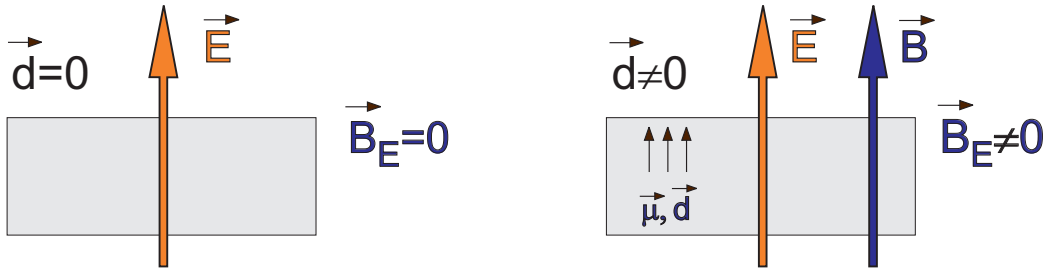}}
\caption{Spins of electrons (atoms) polarized by an electric field induce the magnetic
field $\vec{B}_E$  and change in the magnetic flux $\Phi$ at the
surface of a flat sheet of material}
\end{figure}

If an external magnetic field acts on either a para- or a ferromagnetic material,
the spins in the substance become
polarized due to substance magnetization. 
Therefore, the electric dipole moments appears polarized, too.
This results in the induction of an electric field $\vec{E}_B$ (Fig.3) (see ref. D.DeMille in \cite{1}):
\begin{equation}
{E_B}=4 \pi \rho d P(B),
\label{5}
\end{equation}
where $P$ represents the degree that the spins are polarized in the sample.

\begin{figure}[htbp]
\epsfysize = 3.5 cm \centerline{\epsfbox{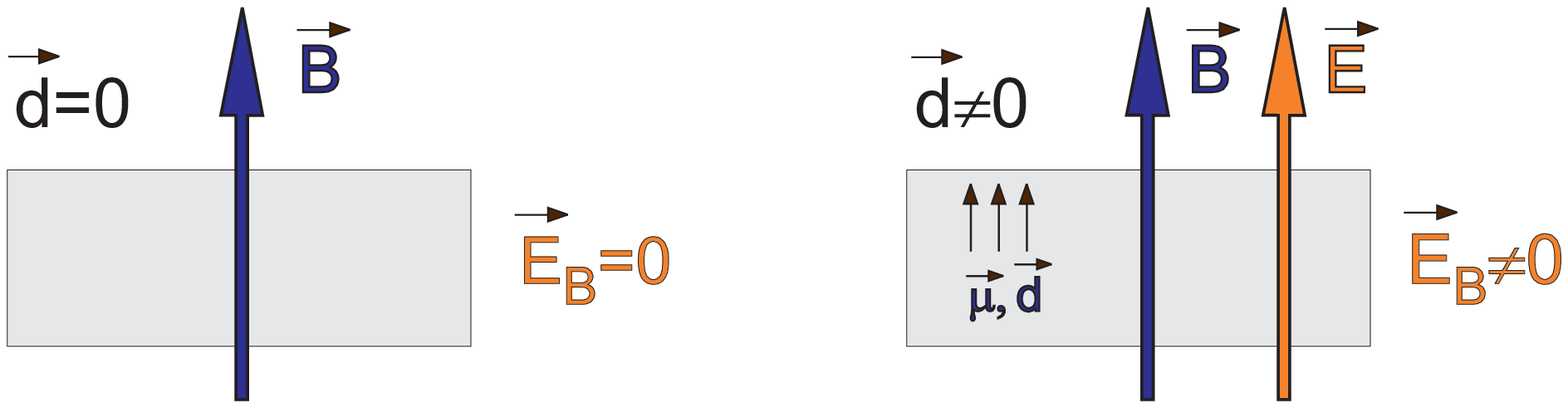}}
\caption{ }
\end{figure}

According to the analysis \cite{1}, 
modern methods for measurement of ${B_E}$ and 
${E_B}$ provide sensitivity for electric dipole moment measurement about
10$^{-32}~e~cm$ and in some cases even 10$^{-35}~e~cm$.

It is important to pay attention to another mechanism of time-reversal violating generation 
of magnetic and electric fields, which have been discussed in \cite{7}.
According to the idea of \cite{7}, 
an induced magnetic moment ${\vec \mu}({\vec E})$ (and, as a result, a magnetic field 
(Fig.4)) of a particle appears due to the action of a field $\vec E$ 
under conditions of violation of P- and T-invariance 
(and similar, an induced electric dipole moment ${\vec d}_B$ 
(an electric field Fig.5) of a particle appears due to the 
action of a field $\vec B$).
This new effect does not depend on temperature.
An effect magnitude is determined by 
a P-odd T-odd tensor polarizability $\beta_{ik}^{T}$
of a particle (atom, molecule, nucleus, neutron, electron and so on).
For an atom (molecule), $\beta_{ik}^{T}$ arises due to P- and T-odd 
interaction of electrons with a nucleus.
For the stationary state of an atom (molecule) $ |N_{0}\rangle$ 
the tensor $\beta_{ik}^{T}$ is as follows:
\begin{equation}
\beta _{ik}^{T}=\sum_{F}\frac{\langle N_{0}|\widehat{d}_{i}|F\rangle \;\langle F|\widehat {\mu}
_{k}|N_{0}\rangle +\langle N_{0}|\widehat {\mu}_{i}|F\rangle \;\langle
F|\widehat {d}_{k}|N_{0}\rangle }{E_{F}-E_{N_{0}}},
\label{bik}
\end{equation}
where $|F\rangle$ is the wave function of a
stationary state of the atom, considering T-odd interaction $V_w^T$,
$E_F$ and $E_{N_0}$ are the energies of the atom (molecule) stationary states,
$\widehat{\overrightarrow{d}}$ and $\widehat{\overrightarrow{\mu}}$
are the operators of electric dipole 
moment and magnetic moment, respectively and $i,k=1,2,3$ correspond to the axes $x,y,z$.

Let us place an atom (molecule) into an electric field $\vec {E}$.
The induced magnetic dipole moment $\vec{\mu} (\vec E)$
appears in this case \cite{7} :
\begin{equation}
{\mu_{i} (\vec E)}=\beta_{ik}^{T} {E}_k,
\end{equation}
The tensor $\beta_{ik}^{T}$ (like any tensor of rank two) can be expanded into scalar, 
symmetric and antisymmetric parts.

The antisymmetric part of the tensor $\beta_{ik}^{T}$ is proportional to
$e_{ikl} J_l$, where $e_{ikl}$ is the totally antisymmetric tensor of rank three.
The symmetric part of the tensor $\beta_{ik}^{T}$ is proportional to the tensor of 
quadrupolarization
$Q_{ik}=\frac{3}{2J(2J-1)}[J_i J_l+J_k J_l-\frac{2}{3} J(J+1)\delta_{ik}]$.
As a result
\begin{equation}
\beta_{ik}^{T}=\beta_{s}^{T} \delta_{ik} + \beta_{v}^{T} e_{ikl} J_l + \beta_{t}^{T} Q_{ik},
\end{equation}
where
$\beta_{s}^{T},\beta_{v}^{T}, \beta_{t}^{T}$ are the scalar, vector and tensor 
P-, T-odd polarizabilities of the particle, respectively.
For a substance with the nonpolarized spins
$Sp~\rho(J){\vec J}=0$ and $Sp~\rho (J) Q_{ik}=0$ (here $\rho (J)$ is the atom 
(molecule) spin density matrix).
As a result for such a substance, $\beta_{ik}^{T}$ appears to be a scalar
$\beta_{ik}^{T}=\delta_{ik} \beta_{s}^{T}$.

Placement of a nonpolarized atom (molecule, nucleus) 
into an electric field induces the magnetic dipole moment $\vec{\mu}_E$:
\begin{equation}
\vec{\mu} (\vec E)=\beta_{s}^{T} \vec{E},
\label{8}
\end{equation}
where $\beta _{s}^{T}=\sum_{F}\frac{\langle N_{0}|\widehat{d}_{z}|F\rangle \;\langle F|\widehat{\mu}
_{z}|N_{0}\rangle +\langle N_{0}|\widehat{\mu} _{z}|F\rangle \;\langle
F|\widehat{d}_{z}|N_{0}\rangle }{E_{F}-E_{N_{0}}}$, 
$\widehat{d}_{z}$ and $\widehat{\mu}_{z}$ are the $z$ components of the operators of the electric dipole 
moment and magnetic moment, respectively, axis $z$ is parallel to the electric field
$\vec{E}$. It should be emphasized that for strong fields (when the distance between atom 
(molecule) levels is comparable with the energy 
of interaction with an electric $\vec E$ (magnetic $\vec B$) field) $\beta_{s}^{T}$ depends on ${\vec E}~({\vec B})$.

\begin{figure}[htbp]
\epsfxsize = 12 cm \centerline{\epsfbox{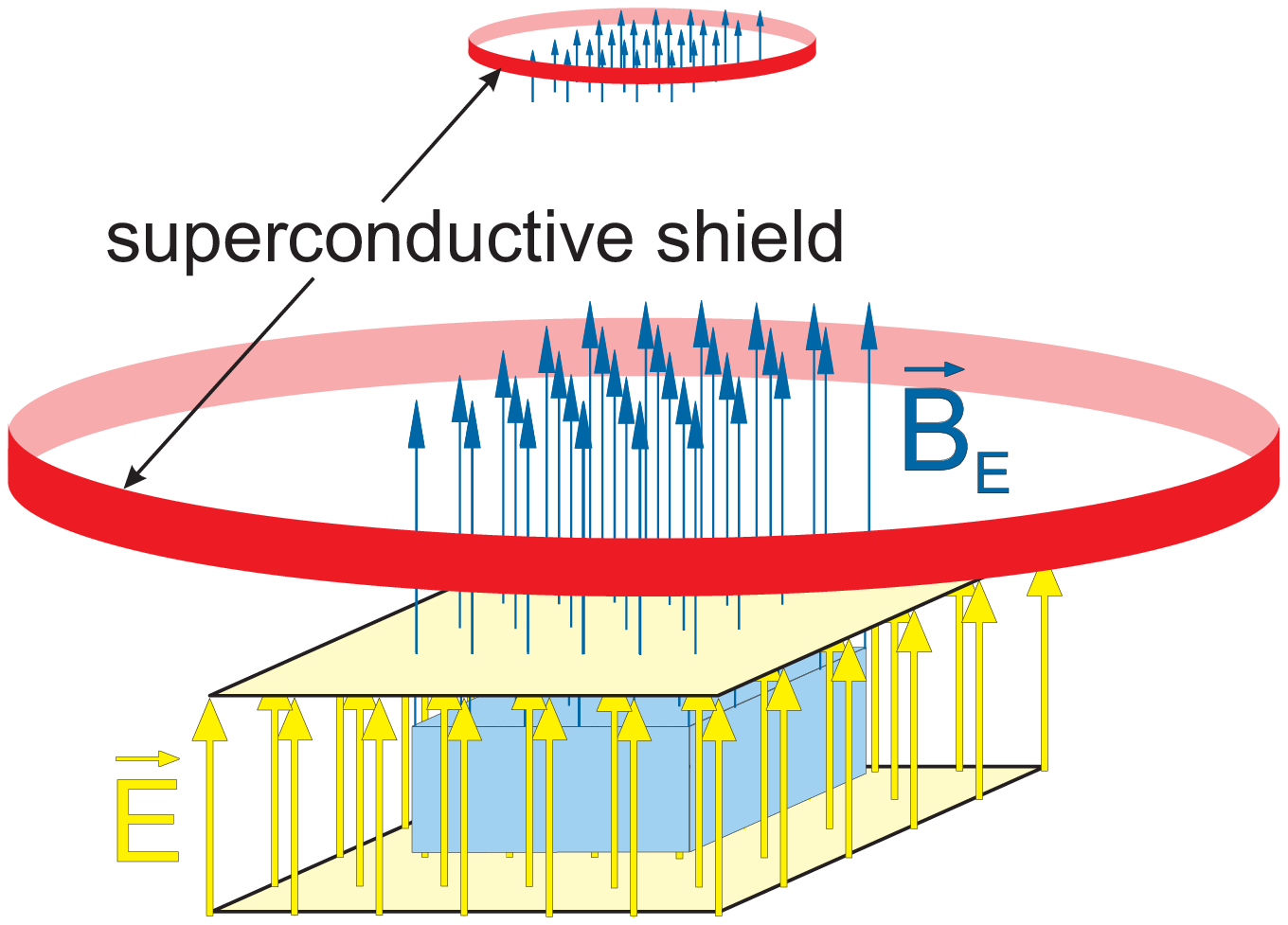}}
\caption{ }
\end{figure}

Weak interaction is much weaker than strong and electromagnetic interactions. Therefore, to find
the wave function $|F\rangle$, the perturbation theory can be applied:
\begin{equation}
|F\rangle=|f\rangle+\sum_{n}
\frac{\langle n|V_{w}^T|f\rangle}
{E_{f}-E_{n}}|n\rangle=
|f\rangle+\sum_{n} \eta_{nf}^T|n\rangle ,
\label{7}
\end{equation}
where $|f\rangle$ is the wave function of an atom in the absence of weak interactions and
the mixing ratio is $\eta_{nf}^T=\frac{\langle n|V_{w}^T|f\rangle}{E_{f}-E_{n}}$.
It should be mentioned that 
for theoretical analysis of $\beta_{s}^T$ in a substance it is necessary
to find a wave function of an excited state of an atom in the substance
which is difficult to do.
%-------------- end of remove

It follows from (\ref{8}) that in a substance placed into
electric field the magnetic field is induced \cite{7}:
\begin{equation}
\vec{B}_{E}^{ind}=4 \pi \rho \beta _{s}^{T} \vec{E}^{*}.
\label{9}
\end{equation}
%end of "remove 1"
%%%%%% "remove 2" should be placed here !!!!!!!!!!!
Vice versa, if an atom (molecule, nucleus) is placed into a magnetic field, 
the induced electric dipole moment $\vec{d}(B)$
appears \cite{7}, 
\begin{equation}
d_{i}(B)=\chi_{ik}^T B_{k},
\label{d_i}
\end{equation} 
where
the tensor polarizability
$\chi_{ik}^T$
is  $\chi_{ik}^T=\beta_{ki}^T$.  
The dipole moment $\vec{d}(B)$ leads to the induction of an electric field 
in the substance:
%\begin{equation}
${E}_{i}^{ind}(B)=4 \pi \rho \beta _{ki}^{T} \vec{B}_{k}^{*} ,
$%\label{10}
%\end{equation}
where $\vec{B^{*}}$ is the local magnetic field, 
acting on the considered particle in the substance.

\begin{figure}[htbp]
\epsfxsize = 9.5 cm \centerline{\epsfbox{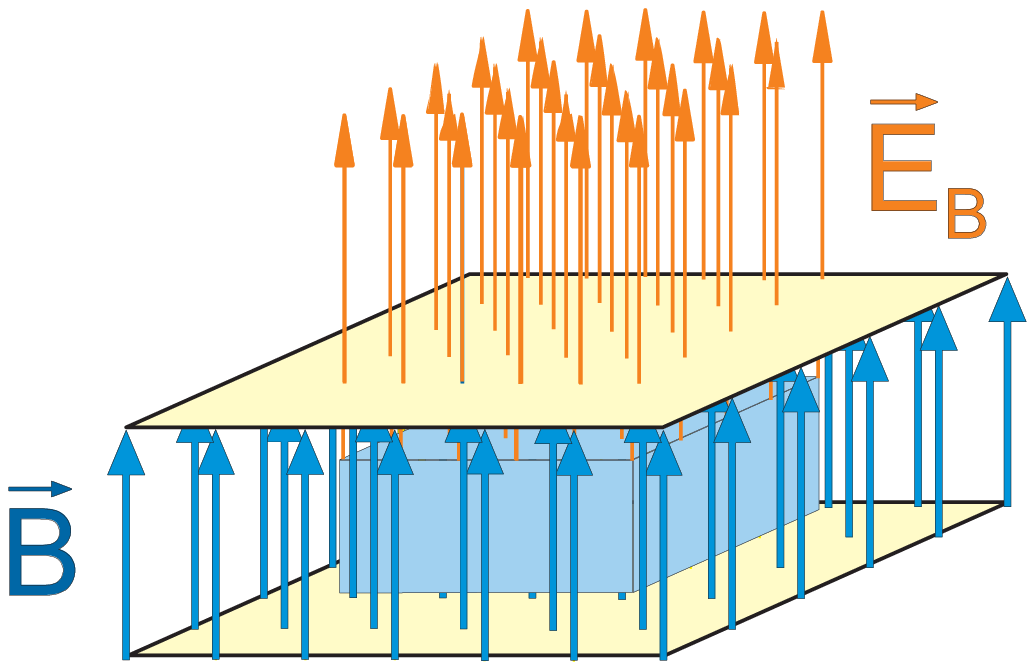}}
\caption{ }
\end{figure}

If an atom is found in a point with the cubic symmetry (or in a liquid),
then $\left\langle Q_{ik}\right\rangle=Sp \rho(J)  Q_{ik}=0$ and
$\left\langle \overrightarrow{J} \right\rangle=Sp \rho(J) \overrightarrow{J} \parallel \overrightarrow{B}^{*}$. 
As a result the terms including $\beta_{v}^T$ and $\beta_{t}^T$ turn to zero
($e_{ikl}\left\langle J_l\right\rangle B_k^{*}=e_{ikl} B_l^{*} B_k^{*}=0$).

As a consequence, in this case
\begin{equation}
{E}_{ind}(B)=4 \pi \rho \beta _{s}^{T} \vec{B}^{*} ,
\label{10_1}
\end{equation}
Hence, analyzing the results of the experiment proposed in
\cite{1}, one should consider that the appearance of the induced magnetic and electric fields is
caused by:
%\begin{enumerate}
%\item 

1. A magnetic field is induced due to interaction of the electric dipole moment of an atom
with an external electric field \cite{1,5} (see (\ref{3}),(\ref{4}))
%\item 

2. A magnetic field is induced due to mechanism \cite{7} (see (\ref{9}))
%\item 

3. A magnetic field appears as a result of polarization (magnetization) of atom magnetic moments
by the local induced magnetic field ${\vec B}_{loc}$ due to interaction $W$ of the magnetic
dipole moment of an atom with this field
\begin{equation}
W=-{\vec{\mu}_a} {\vec B}_{loc}.
\end{equation}
The local field ${\vec B}_{loc}$ is the sum of two contributions:
\begin{equation}
{\vec B}_{loc}={\vec B}_{E~loc}+{\vec B}_{loc}^{ind},
\end{equation}
where the field ${\vec B}_{E~loc}$ is the local magnetic field acting on an atom from the polarized
(by mechanism [1,5], see (3),(4)) magnetic moments of the other atoms of the sample. 
This field depends on temperature and its contribution could be neglected for those temperature values,
which provide $\chi \ll 1$. But for temperature $T<1K$ 
the susceptibility $\chi \sim \frac{1}{T}$ becomes comparable with $1$
and higher, and the energy of interaction of two magnetic dipoles for neighbour atoms 
occurs of order of $k_{B} T$ and greater. Thus, in this case, the collective effects, well-known in the 
theory of phase transition
in magnetism, should be taken into account while considering magnetization by an electric field.

The field ${\vec B}_{loc}^{ind}={\vec B}_{1~loc}^{ind}+{\vec B}_{2~loc}^{ind}$ does not depend on temperature.

The field ${\vec B}_{1~loc}^{ind}=\chi_{1~loc}^T {\vec E}^{*}$ is the local magnetic field produced 
in the point of the considered atom location
by the magnetic moments 
of atoms of the substance (except for the considered atom)
induced by the aid of mechanism \cite{7}
(see (6),(8));
$\chi_{1~loc}^T \sim \rho \beta _{s}^{T}$ is the local P-,T-odd susceptibility of the substance
In general case $\chi^T_{1~loc}$ is a tensor, but if atom surrounding
posseses cubic symmetry, then the principle contribution to this tensor 
is made by its scalar part 
($\chi^T_{1~loc}$ depends on the substance density and sample shape: 
for sphere $\chi_{1~loc}^T=\frac{8 \pi}{3} \rho \beta _{s}^{T}$;
for cylinder $\chi_{1~loc}^T={4 \pi} \rho \beta _{s}^{T}$ ).

The field
${\vec B}_{2~loc}^{ind}$ is the self-induced magnetic field of the considered atom. 
% ---- style ------
The
magnetic moment 
(T-odd current) of the atom induced by an electric field acting on 
the atom due to mechanism [7,8] causes 
appearance of the magnetic field inside the atom:
\begin{equation}
{\vec H}_{E}^{T}({\vec r})=rot~{\vec A}_{E}^{T}({\vec r}) 
\end{equation}
with the vector potential 
${\vec A}_{E}^{T}({\vec r})=\frac{1}{c} \int \frac
{j_{E}^{T}({\vec r}^{\prime})}{\left|{\vec r}-{\vec r}^{\prime})\right|} d^3 r^{\prime}$,
$j_{E}^{T}({\vec r}^{\prime})$ is the T-odd part of the transition current density operator
for an atom (molecule) placed in an electric field \cite{12}
(it is calculated by the use of wavefunctions (10)).
The magnetic interaction Hamiltonian 
of an atom with the field ${\vec A}_{E}^{T}$
can be expressed as \cite{new_lanl}:
\begin{equation}
W_{2~loc}=-\frac{1}{2c} \int 
(
{\vec j}({\vec r}) {\vec A}^{T}({\vec r})+
{\vec A}^{T}({\vec r}) {\vec j}({\vec r}) 
)
d^3 r,
\end{equation}
where ${\vec j}({\vec r})$ is the atom transition current density operator calculated with the atom wavefunction 
without consideration of P-,T-odd interactions,
%\begin{equation}
${\vec j}({\vec r})=c~rot{\vec {\mu}}({\vec r})$
%\end{equation} 
and ${\vec {\mu}}({\vec r})$ is the operator 
of the atom magnetic moment density.
If the atom is found in a point with the cubic symmetry (or in a liquid)
then we may omit contributions from atom multipoles.

In this case the above expression can be written as:
\begin{equation}
W_{2~loc}=-\overrightarrow{\mu}_a \overrightarrow{B}_{ind}
=-\chi_a^T \overrightarrow{\mu}_a \overrightarrow{E}_{loc}^{*},
\label{W2loc}
\end{equation}
%%%%%%%%%%%%%%%%%%%%
where  
${\vec {\mu}}_a={\mu}_a \frac{{\vec J}}{J}$
and $\chi_{at}^T$ is the T-odd atom susceptibility, which does not depend on the substance density 
and sample shape,
$\chi_{at}^T \sim \beta _{s}^{T} \frac{1}{a^3}$ 
(here $a$ is the typical radius of distribution density  of the magnetic
moment induced by an electric field in the atom \cite{7}).

As a result one obtains:
\begin{eqnarray}
{\vec B}_{loc}^{ind}=(\chi_{1~loc}^T + \chi_{at}^T) {\vec E}_{loc}^{*}=\chi_{loc(subst)}^T {\vec E}_{loc}^{*},\\
\chi_{loc(subst)}^T=\chi_{1~loc}^T + \chi_{at}^T.
\end{eqnarray}

%\end{enumerate}

The interaction \ref{2} of the magnetic moment of an atom with the induced magnetic field 
causes the appearance of the magnetic field due to different
population of magnetic levels of the atom in the field ${\vec B}_{loc}$
in thermal equilibrium
\begin{equation}
{\vec B}^{\prime}_{ind}=
4 \pi \chi {\vec B}_{loc} \approx
4 \pi \chi {\vec B}_{loc}^{ind}=
4 \pi \frac{\rho \mu_{a}^{2}}{3 k_{B} T} \chi_{loc(subst)}^T {\vec E}^{*},
\end{equation}
the field ${\vec B}_{loc}={\vec B}_{E~loc}+{\vec B}_{loc}^{ind}$, but ${\vec B}_{E~loc}$ 
contribution could be neglected for those temperature values,
which provide $\chi \ll 1$, and it is omitted here.

Therefore, the flux $\Delta \Phi$, which is going to be measured in the 
experiment proposed in
\cite{1} should be written as:
\begin{eqnarray}
& & \Delta \Phi  =  A B_{E}= 4 \pi A (\chi \frac{d}{\mu_a}+ \rho \beta _{s}^{T}+ \chi \chi_{loc(subst)}^T)E^{*}
= \nonumber \\ & & =  
4 \pi A [ \chi (\frac{d+\mu_a \chi_{at}^T}{\mu_a}+\chi^T_{1~loc})+\rho \beta_{s}^T]E^{*},
\label{11} \\
& & {\vec B}_E = 
4 \pi [\chi (\frac{d+\mu_a \chi_{at}^T}{\mu_a}+\chi_{1~loc}^T) +\rho \beta _{s}^{T}]E^{*},
\label{12}
\end{eqnarray}
where $\chi=\frac{\rho \mu_{a}^2}{3 k_{B}T}$.

\begin{figure}[htbp]
\epsfxsize = 14.5 cm \centerline{\epsfbox{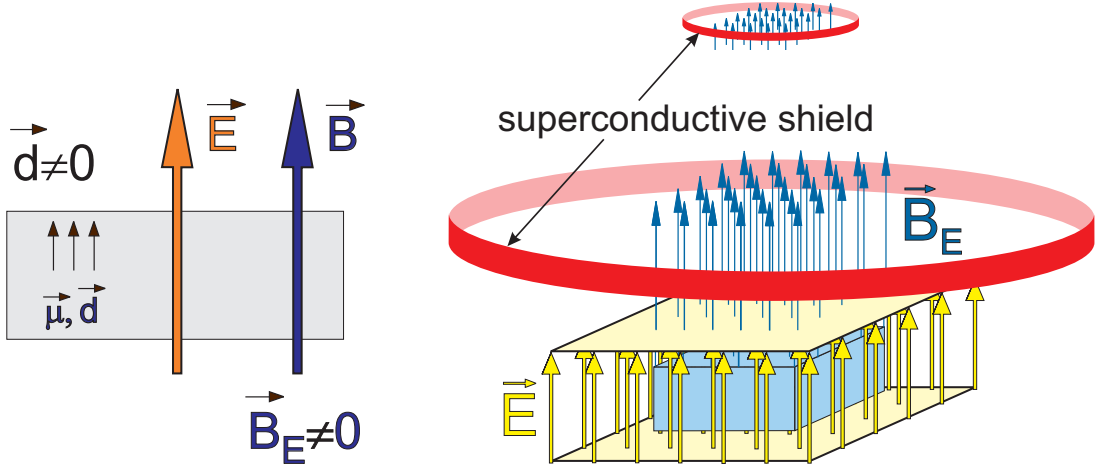}}
\caption{The magnetic field that is observed in an experiment is the sum 
of magnetic fields produced by the polarized spins of particles and 
induced by the external electric field ($\sim \beta^T$)}
\end{figure}

It should be noted that the quantity $d+\mu_a \chi_{at}^T$ is the electric dipole moment 
of an atom available for measurement in 
conventional EDM experiments studying atom (molecule) spin precession in an external
electric field.
It is well known that the atom EDM is arisen from several mechanisms:
the contribution proportional to $d_e$, the contribution due to T-,P-odd interaction
of atom electrons with nuclear nucleons (dependent and independent on nuclear spin) \cite{9}
% insert
($d$  also contains contribution $\sim \beta_s^T \frac{\mu_a}{a^3}$ induced according to 
(\ref{d_i}) by the magnetic field  $\sim \frac{\mu_a}{a^3}$, which is produced inside the paramagnetic 
atom by its electrons).

According to (\ref{W2loc},\ref{12}) 
there is one more addition to the EDM $ \sim \mu_a \chi_{at}^T$.

Note that the contribution to the atom EDM proportional to the nucleus spin $I$
is equal to zero if the sample temperature is high ($\left\langle I\right\rangle=0$ at high temperatures).

Let us consider now the experiment to detect the electric dipole moment of the electron
by means of measurement of the electric field \cite{1} (see (\ref{5})). 
In this case we also should take into consideration
the effect \cite{7} of EDM induction by the magnetic field (\ref{d_i}).

\begin{figure}[htbp]
\epsfxsize = 14.5 cm \centerline{\epsfbox{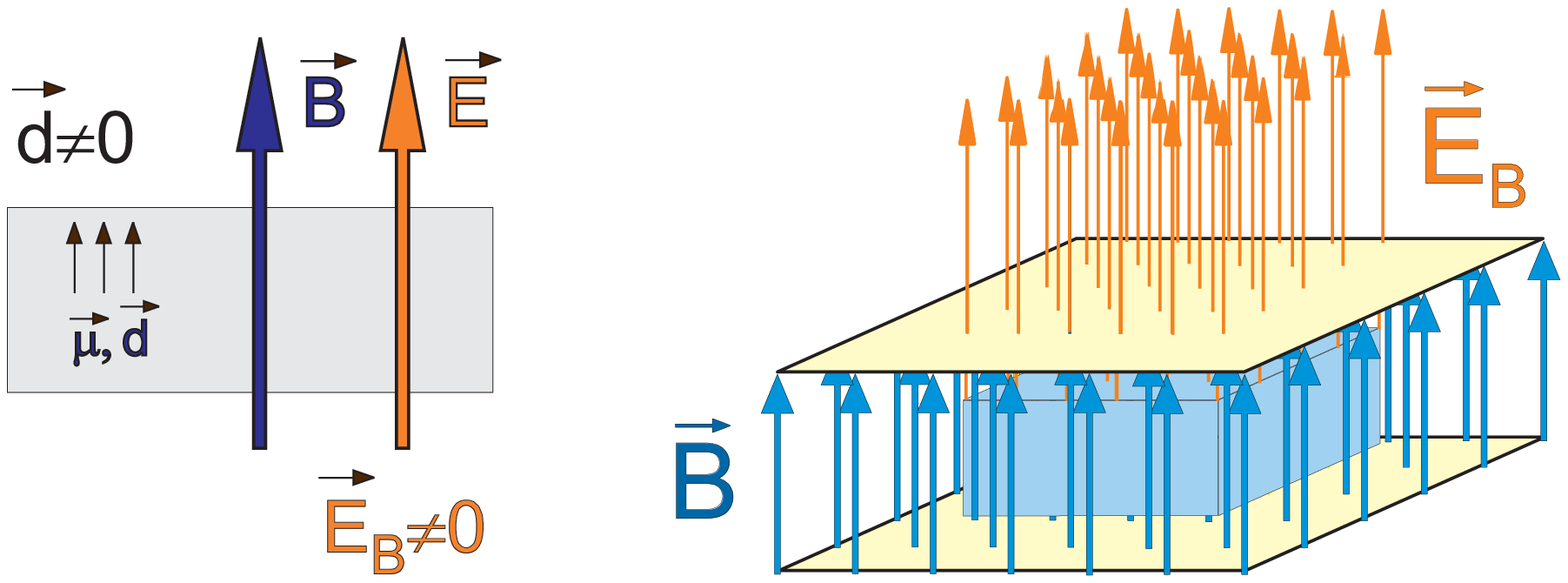}}
\caption{ }
\end{figure}

Thus, the electric field measured in the experiment \cite{1} is as follows
(see (9),(10), Fig.7):
\begin{equation}
{E_B}=4 \pi \rho (d_a P(B)+  \beta _{s}^{T} B^{*}),
\label{13}
\end{equation}
$d_a$ is the atom EDM containing contribution $\sim \beta_s^T \frac{\mu_a}{a^3}$ induced according to 
(\ref{d_i}) by the magnetic field  $\sim \frac{\mu_a}{a^3}$, which is produced inside the paramagnetic 
atom by its electrons.

So, measurement of $\Delta \Phi$ and ${E_B}$ 
provides knowledge about the atom EDM  
and $\beta _{s}^{T}$.
To distinguish these contributions one should consider the fact that 
$\chi$ and $P(B)$ depend on temperature, 
while $\beta _{s}^{T}$ does not.
Therefore, studying $B_E$ and $E_B$ dependence on temperature allows one to
evaluate contributions from the EDM and $\beta _{s}^{T}$ to the measured effect.

It should be particularly emphasized that $\Delta \Phi$ and ${E_B}$ differs from zero even
when the electron EDM $d_e$ is equal to zero.

It should be also emphasized that the polarizability $\beta _{s}^{T}$
differs from zero even 
for atoms with the zero spin, for which EDM is absent.
In this case 
(as well as for high temperatures when the average atom spin $\left\langle {\vec J}\right\rangle=0$) 
only the effect \cite{7}, described by the terms 
(\ref{12},\ref{13}), proportional to $\beta _{s}^{T}$, 
contributes to the induced electric and magnetic fields.
If the substance consists of several types of atoms, then
their contribution to the induced field is expressed as a sum of contributions from different atoms:
\begin{equation}
\vec{B}_E=4 \pi \rho \sum\limits_{n}c_{n}\beta _{ns}^{T} \vec{E}_n^{*}, 
\vec{E}_B=4 \pi \rho \sum\limits_{n}c_{n}\beta _{ns}^{T} \vec{B}_{n}^{*},
\label{*}
\end{equation}
where $c_n$ is the concentration of atoms of the type $n$, $\vec{E}_n^{*}$
and $\vec{B}_n^{*}$ are the local fields acting on atoms of the type $n$.

Now let us consider what information about constants of T-,P-odd interaction of an electron with
a nucleus can be obtained from studying the effect \cite{7} of time-reversal violating generation of
fields $\vec{E}_B$ and $\vec{B}_{E}$ (describing by (\ref{*})).

According to \cite{1} we can expect a magnetic induction sensitivity about
$3 \times 10^{-15}~G/ \sqrt{Hz}$. In ten days of averaging the  sensitivity
is $\sim 10^{-18}~G$. This leads to the sensitivity for $d_e$ measurement 
of about $10^{-32}~e~cm$ \cite{1}. 

Such sensitivity of magnetic induction measurement provides for polarizability $\beta_s^T$ measurement
the sensitivity $\beta_s^T=\frac{B_E}{4 \pi \rho E^{*}} \sim 10^{-43}$ cm$^3$ (it is supposed that
$\rho \approx 2 \div 3 \cdot 10^{22}$ and $E^{*} \sim 10 ~\frac{kV}{cm}$, for example, for liquid and solid
Xe $\rho \approx 2 \cdot 10^{22}$).

Let us consider now the possibilities given by the experiment studying the electric field, which is induced by
a magnetic field \cite{7} for measurement of the polarizability $\beta_s^T$.
Analysis \cite{1} shows that existing methods of the electric field measurement allow to measure
electric fields 
$E \sim 10^{-13} \div 10^{-14} \frac{V}{cm} \sim 3 \cdot 10^{-16} \div 3 \cdot 10^{-17}$ CGSE 
in ten days operation. Therefore, for $\beta_s^T=\frac{E_B}{4 \pi \rho B}$ we can
get the estimation $\beta_s^T \approx 10^{-43} \div 10^{-45}$ cm$^3$ 
($\rho \approx 2 \cdot 10^{22}$, $B \approx 10^{4} \div 5 \cdot 10^{4}$ Gauss).

The obtained evaluation for $\beta_s^T$ ($\beta_s^T \sim 10^{-43} \div 10^{-45}$ cm$^3$) and the expressions
(\ref{bik},\ref{7}) allow us to evaluate the mixing ratio $\eta_{nf}^T$.
Recall that conventional T-,P-even polarizability of an atom can be expressed similar
(\ref{bik}) with replacement of the matrix element $\mu$ by the matrix element $d$
(and $\eta_{nf}^T=0$). Therefore, we can estimate $\beta_s^T \sim \beta_s \alpha \eta_T$,
where $\beta_s$ is the conventional T-,P-even polarizability of the atom, $\alpha=\frac{1}{137}$
is the fine structure constant and $\eta_T$ is some average value for the coefficient of mixing
of opposite parity levels $\eta_{nf}^T$. The estimation for $\eta_T$ follows from the above:
$\eta_T \sim \frac{\beta_s^T}{\beta_s \alpha} \approx 10^{-17} \div 10^{-19}$ (the atom polarizability
is of the order $\beta_s \sim 10^{-24}$ cm$^3$, for example, for Xe according to \cite{Kozlov}
$\beta_s=2.7 \cdot 10^{-24}$ cm$^3$).

Two types of T-,P-odd interactions contribute to the constant of mixing of 
opposite parity levels $\eta_{nf}^T$: the interaction of the electron EDM with the
coulomb field of the nucleus and the T-,P-odd interaction of the electron with 
the nucleus nucleons.

The spin-independent part of the T-,P-odd interaction of an electron with nucleons is described
by two constants \cite{9}: $k_{1p}$ describes interaction with protons and
$k_{1n}$ describes interaction with neutrons. For example, 
calculation adduced in \cite{9} for $Cs$ provides 
$\eta_T=3.7 \cdot 10^{-11} (0.41 k_{1p}+0.59 k_{1n})$, 
where the sum $(0.41 k_{1p}+0.59 k_{1n})<5 \cdot 10^{-4}$.
Therefore, for Cs $\eta_T \lesssim 10^{-14}$.
The same limits for the sum $k_{1p}$ and $k_{1n}$ were obtained from
EDM measurements for $^{129}Xe$ \cite{9} $(0.4 k_{1p}+0.6 k_{1n}) \lesssim 10^{-4}$.

Let us note that in the experiment \cite{Cs} planned to measure the EDM of $Cs$ atoms
trapped in optical lattices is expected to obtain sensitivity of Cs EDM measurement of about 
$d_{Cs} \approx 3 \cdot 10^{-25}$ $e \cdot cm$. This value of the atom EDM provides to improve
estimation for $\eta_T$ and to get limits $\eta_T \lesssim 10^{-16}$ and the sum of 
$k_{1p}$, $k_{1n}$ $\lesssim 5 \cdot 10^{-6}$.

As it was shown above the experiments studying $\beta_s^T$ provide for mixing
coefficient the value $\eta_T \lesssim 10^{-17} \div 10^{-19}$. Therefore, these experiments give hope 
to reduce (three orders) the estimations for the sum $k_{1p}$, $k_{1n}$ (it is expected to be
$\lesssim 10^{-7} \div 10^{-9}$). 
This value for $k_{1p}$,  $k_{1n}$ is significantly lower than
the limitation 
%% to check
which could be obtained from the 
results of measurements
of atom dipole moment that have been done earlier (see, for example, \cite{9})
and from the proposed experiment \cite{Cs}.

Let us consider now what limits for the electron EDM $d_e$ can be obtained
from study of the effect \cite{7}. To estimate contribution from $d_e$
to polarizability $\beta_s^T \sim \beta_s \alpha \eta_T$ let us use 
connection of mixing coefficient $\eta_T^e$ (caused by the electron EDM $d_e$)
with the atom EDM $d_A$ induced by the electron EDM $d_e$: 
$d_A \sim e a \eta_T^e \sim R d_e$ \cite{9},
$R$ is the atomic EDM enhanced factor. Therefore,
$\eta_T^e \sim \frac{R d_e}{e a}$. Using the
mentioned estimations for $\beta_s^T$ and $\eta_T^e$
we obtain $d_e \sim \frac{a \beta_s^T}{\beta_s \alpha R}~e \cdot cm$
i.e. $d_e \sim 10^{-27} \div 10^{-30}~ e \cdot cm$
($\beta_s^T \approx 10^{-43} \div 10^{-45}~cm^3$, $R \approx 10^2 \div 5 \cdot 10^2$).
Let us remind that the upper limit $d_e \lesssim 1.6 \cdot 10^{-27}~e \cdot cm$ 
follows from the experiments with Tl \cite{3}.

Thus, experimental study of the effect \cite{7} 
provides to get more (three or four orders)
strict limits for constants, which describe T-,P-odd interactions of an electron with nucleons and for $d_e$.
Such improvement 
of estimations 
constrains theories beyond the Standard Model.
%aggravates the limitations for theories describing T-,P-odd interactions.
To study effect \cite{7} one could use different atoms (molecules) and substances
(for example, ferroelectric crystals providing
very high electric fields for heavy atoms) and this study does not require 
target cooling to ultralow temperatures (1 K and lower).
Spins of atoms (molecules) can be nonpolarized or atoms (molecules) can be spinless.

\end{document}